\documentclass[sigconf]{acmart}

\usepackage{booktabs} 
\usepackage{siunitx}

\begin{document}
\title{Pible: Battery-Free Mote for Perpetual Indoor BLE Applications}

\author{Francesco Fraternali}
\orcid{1234-5678-9012-3456}
\affiliation{%
  \institution{University of California, San Diego}
  }
  \email{frfrater@ucsd.edu}
\author{Bharathan Balaji}
\affiliation{%
  \institution{University of California, Los Angeles}
}
\email{bbalaji@ucla.edu }

\author{Yuvraj Agarwal}
\affiliation{%
  \institution{Carnegie Mellon University}
  }
\email{yuvraj@cs.cmu.edu }  
  
\author{Luca Benini}
\affiliation{%
  \institution{University of Bologna - ETH Zurich}
  }
\email{luca.benini@unibo.it }  
  
\author{Rajesh Gupta}
\affiliation{%
  \institution{University of California, San Diego}
}
\email{gupta@eng.ucsd.edu }

\renewcommand{\shortauthors}{F. Fraternali et al.}

\begin{abstract}
Smart building applications require a large-scale deployment of sensors distributed across the environment. Recent innovations in smart environments are driven by wireless networked sensors as they are easy to deploy.
However, replacing these batteries at scale is a non-trivial, labor-intensive task. Energy harvesting has emerged as a potential solution to avoid battery replacement but requires compromises such as application specific design, simplified communication protocol or reduced quality of service. We explore the design space of battery-free sensor nodes using commercial off the shelf components, and present Pible: a Perpetual Indoor BLE sensor node that leverages ambient light and can support numerous smart building applications.
We analyze node-lifetime, quality of service and light availability trade-offs and present a predictive algorithm that adapts to changing lighting conditions to maximize node lifetime and application quality of service. Using a 20 node, 15-day deployment in a real building under varying lighting conditions, we show feasible applications that can be implemented using Pible and the boundary conditions under which they can fail.  
\end{abstract}

%
%



\acmYear{2018}\copyrightyear{2018}
\setcopyright{acmcopyright}
\acmConference[BuildSys '18]{BuildSys '18: Conference on Systems for Built Environments}{November 7--8, 2018}{Shenzen, China}
\acmBooktitle{BuildSys '18: Conference on Systems for Built Environments, November 7--8, 2018, Shenzen, China}
\acmPrice{15.00}
\acmDOI{10.1145/3276774.3276785}
\acmISBN{978-1-4503-5951-1/18/11}

\maketitle

\section{Introduction}
Buildings are integrated with thousands of sensors: temperature sensors provide feedback to HVAC systems, smoke sensors provide fire safety, etc. 
These sensing systems are designed for wired communication and power during the design of the building itself. 
Even a minor change requires a domain expertise in the building's wiring infrastructure 
and can be prohibitively expensive, e.g., 
\$2500 \cite{link:retrofitting-cost}. 
Wireless sensors have emerged as the answer to this 
problem. With low power and 
communication protocols (i.e ZigBee, 6LowPAN), 
wireless sensors can be deployed with a multi-year battery lifetime 
and be used for applications such as 
occupancy based control~\cite{paper:yuvraj_1}. 
But these nodes are powered by batteries that require periodic manual replacement.
As we scale to large deployments, the manual replacement of batteries becomes a bottleneck. 
Battery replacements can be mitigated using energy harvesting, e.g., DoubleDip measures water flow by powering itself using temperature difference~\cite{paper:DoubleDip}. 
However, there are limited commercial devices that use indoor energy harvesting solutions. We highlight 3 limitations that inhibit adoption: : (i) they are designed for specific applications, (ii) they do not support standard protocols, (iii) the application quality of service (QoS) is inadequate.

In this paper, we analyze the feasibility of overcoming these limitations using commercial off the shelf components. We explore the design space of a generic energy harvesting sensor node for indoor monitoring applications with the objective of perpetual operations \cite{paper:campbell_1}. 
We designed and built \emph{Pible}, a Perpetual Indoor BLE sensor node. 
We show trade-offs between QoS, lifetime and harvested energy that enables our prototype sensor node to work in different indoor lighting conditions. 
We introduce hardware solutions to increase charging efficiency and overcome cold-start operations that limit 
usability. Finally, we propose a local sensor-node power management solution that maximizes the application-QoS and node-lifetime.
We evaluate Pible by deploying twenty nodes in five different lighting conditions, for a general set of building sensing applications such as periodic sensor measurements, e.g. temperature, and event-driven sensors, e.g. PIR. We conducted a 15-day experiment in which we demonstrate continuous operation for all the different applications and on every light condition tested. Results show that Pible is able to broadcast sensor data at an average period of 94 seconds with an average luminance of 235 lux per day. 
\vspace{-2mm}

\section{Background and Related Work}
\label{sub:relatedwork}
Existing works in energy harvesting can be categorized into:

\textit{(i) Application Specific:}
Application specific indoor systems have been proposed that use energy harvesting~\cite{paper:DoubleDip}. 
The company EnOcean~\cite{link:enocean} produces battery-less systems 
but their design is application specific. We use a generic design suitable for most applications. 

\textit{ (ii) Energy Harvesting + Rechargeable Battery}: energy harvesting can be used to extend batteries lifetime~\cite{paper:DoubleDip}. 
However, the life of rechargeable batteries is limited to a few hundred cycles~\cite{link:supercap_2}. 
Hence, recent works and our Pible node exploit super-capacitors as they can support up to a million charging cycles.


\textit{(iv) Communication protocol}:
Prior energy harvesting works do not support standard protocols since they require more energy
\cite{paper:campbell_2}. To facilitate integration with existing technologies, we set this as a requirement. 
Works in \cite{paper:EH-ZigBee-6406192,link:enocean,paper:SorberFlickerHester:2017:FRP:3131672.3131674} adopt standard protocols but either lack in perpetual operations or are application specific. 

\textit{(v) Power Management:} Work in \cite{paper:Mani-power-management} adopt a power management 
to achieve energy-neutral operations. We use a simple application-specific look-up table, and we show that Pible achieves perpetual operation under varying lighting conditions. 

\textit{(vi) Intermittent:} Campbell et al.~\cite{paper:campbell_1, paper:campbell_2} designed an indoor sensor-node 
that stores only the amount of energy needed to read and transmit a single data packet: the system is continuously working when light is available 
but it stops during dark periods. 
Results show that door-open detection achieves only 66\% accuracy. 

Table \ref{tab:Introduction}, shows the gaps between related work and battery motes and compare them with Pible. 
 We categorize them into 3 buckets: (i) battery powered systems, (ii) battery and energy harvesting systems and (iii) systems using only harvested energy. 
Battery powered systems achieve high QoS for the applications at the expense of a periodic battery replacement requirement. Adding an energy harvesting solution to a battery powered device increases the lifetime but the batteries still need to be replaced. Prior works in energy harvesting nodes promise no battery replacement but compromise the QoS, do not support standard protocols or are application specific. Flicker~\cite{paper:SorberFlickerHester:2017:FRP:3131672.3131674} does provide these functionalities, but has not been evaluated indoors where we have limited light availability.



\vspace{-3mm}
\begin{table}[ht]
\centering
\footnotesize
\caption{Pible comparison with State of the Art Solutions}
\vspace{-3mm}
\label{tab:Introduction}
    \begin{tabular}{c |c |c|c|c|c}
    \toprule
    Platform				& 	Bat/			& Replace &   Quality 		& 	Standard			& Application\\
					&  	Har					& Battery 	&  		of	&  	 Commun	& \\
									&  			Power			& 	&  			Service	&  	 Protocol	& \\

    \midrule
    Synergy\cite{paper:yuvraj_1} 	&		Bat					&    No 	&	$\sim$GT		&  Zigbee						& Occupancy\\
    \midrule
    
    Trinity	\cite{paper:trinity} 		& 	Bat-Har				& 	 No		&  	$\sim$GT 	& 	Zigbee			& Air-Flow\\
    
    DoubleDip\cite{paper:DoubleDip}	& 		Bat-Har			&   No 		&   98-65\%		&		No						& Water-Flow\\
        \midrule
    Enocean	\cite{link:enocean}		& 		Har					&   Yes		&   NA		&		BLE						& Specific\\
    
    Buzz \cite{paper:campbell_2}		& 		Har					&   Yes 	&   66\%		&		No						& Door, Light\\
    
    Flicker\cite{paper:SorberFlickerHester:2017:FRP:3131672.3131674}		& 		Har					&   Yes 	&   NA		&		BLE						& Periodic Sense\\
    
    

    \midrule
    
    Pible						&		Har				&  Yes		&  $\sim$GT 	& 		BLE						&  Sense/Event\\
  			&							& 		&  	& 							& Occupancy\\
    \midrule
    \multicolumn{6}{ c }{GT = Ground Truth; Bat = Battery; Har = Harvesting}\\\hline
  \end{tabular}
  
\end{table}

\vspace{-5mm}

\section{Design-Space and Architecture}
\label{sec:Architecture}
We select a common set of applications 
in smart buildings 
and assess their power budgets. We select commercial off the shelf components to support their QoS requirements while ensuring perpetual operation on typical indoor environment without a battery. 
\vspace{-2mm}
\subsection{Indoor Monitoring Applications}
We categorize indoor applications as periodic or event-driven: 

\textbf{Sensing Environmental Conditions:}
\textit{(i) 1 Sensor}: We test if Pible can operate perpetually for applications with minimal energy budget such as sensing of 1 sensor. Results 
can be extended to sensors with similar power budget.
\textit{(ii) Multiple Sensors}:
We extend the power budget of motes that monitor different sensors at once. 
 
\textbf{Occupancy Detection:} 
\textit{(i) PIR Motion Sensor}, \textit{Door Sensor}~\cite{paper:yuvraj_1}.
\textit{(ii) Bluetooth Low Energy (BLE) Beacons}~\cite{paper:Cosero}.

Table \ref{tab:Energy} reports the power consumption of Pible's main components and operations using a super-capacitor storage element charged at 3V. For the sensing operations (i.e. Read Temperature), the power includes the reading and BLE transmission. 

\vspace{-2mm}
\begin{table}[ht]
\caption{Power consumption of Pible key operations executed with a super-capacitor charged at 3V. The values reported are averaged on a one minute execution.}
\centering
\vspace{-3mm}
\footnotesize
\label{tab:Energy}
    \begin{tabular}{c c c c}
    \toprule
    Operation & Power[\si\micro W] & Operation & Power[\si\micro W]\\ 
	\midrule
	\small{MCU-Sleep} & \small{19} & \small{PIR Detection} & \small{32} \\
	\small{Read-Hum} & \small{51} & \small{Advertising-5s} & \small{69} \\
	\small{Read-Temp} & \small{54} & \small{Advertising-2s} & \small{86} \\
    \small{Read-Bar} & \small{54} & \small{Advertising-1s} & \small{106}\\ 
    \small{Read-Light} & \small{47} & \small{Advertising-500ms} & \small{171}\\ 
    \small{MCU+PIR Sleep} & \small{22}& \small{Advertising-100ms} & \small{648}\\
    \bottomrule
    \end{tabular}
    \vspace{-5mm}
\end{table}

\vspace{-1mm}
\subsection{Pible-Architecture}
We use a general energy harvesting architecture \cite{paper:EH_review, Scaling_my}: an energy harvester transfers power to a storage element through an energy management board (EMB). Once the energy stored reaches a usable level, the EMB powers the micro-controller that starts its operations. 

\vspace{-2mm}
\subsubsection{Platform System on Chip, Antenna and Sensors}
We select the TI CC2650 chip that supports multiple communication protocols (e.g. 6LoWPAN, BLE) and consumes 1 \si\micro A in standby mode.  
We equip our board with temperature, light, humidity, pressure, reed switch, accelerometer, gyroscope, and a PIR motion sensor.
\vspace{-2mm}
\subsubsection{Energy Storage}
\label{battery}
batteries have a short cycle-life of 1000 recharges. To increase lifetime, we adopt super-capacitors as they support up to 1 million recharges~\cite{link:supercap_2}. However, 
super-capacitors' choice is not trivial since:
\textit{(i)} Energy stored 
drops linearly on discharge~\cite{link:supercap_2}.
\textit{(ii)} super-capacitor size is proportional to charging time using a constant charging current. 
\textit{(iii)} Large super-capacitors increase node dimensions leading to packaging and aesthetic issues.
\textit{(iv)} The leakage current increases with size.
We tested 3 super-capacitors with capacitance of 0.22F, 0.44F and 1F at 3.6V. 
\vspace{-2mm}
\subsubsection{Energy Harvester}
We use solar light since it has high power density 
w.r.t other energy sources~\cite{paper:EH_review} and it is available in buildings. To match our worst case scenario (Table \ref{tab:Energy}), we use the indoor solar panel AM-1454 that harvests 
71 \si\micro W with 300lux. 

\vspace{-2mm}
\subsubsection{Energy Management Board (EMB)}
\begin{figure}[ht]
\vspace{-2mm}
	\centering
	\includegraphics[width=\linewidth]{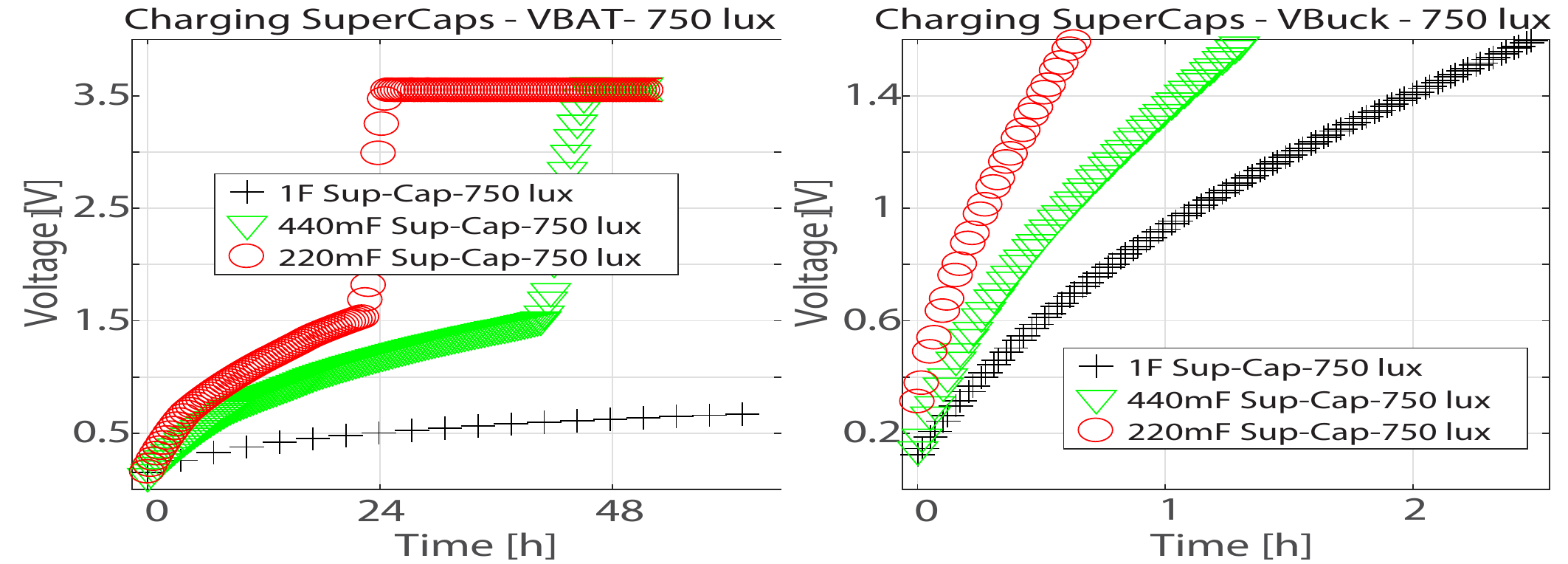}
	\vspace{-3mm}
	\caption{Charging different Super-Capacitors Size: Left- 750 lux with VBAT, Right: 750 lux with VBuck}
	\label{fig:ChargeColdStart}
	\vspace{-2mm}
\end{figure}

\begin{figure}[ht]
\vspace{-2mm}
	\centering
	\includegraphics[width=0.7\linewidth]{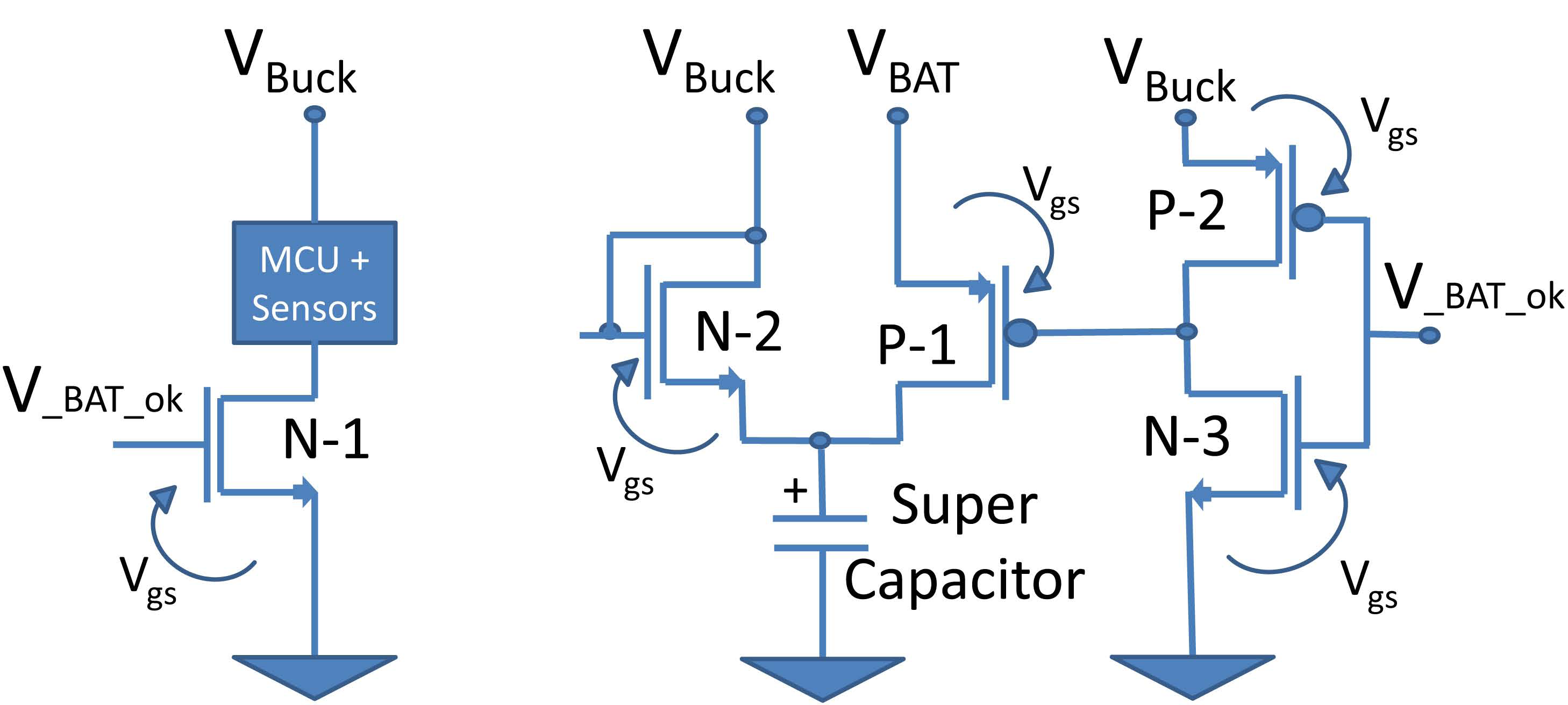}
	\vspace{-3mm}
	\caption{Our circuitry to switch between V$_{BAT}$ and V$_{Buck}$.}
	\label{fig:CircuitMOS}
	\vspace{-3mm}
\end{figure}

For our design, we select the BQ25570 from TI since it includes two programmable DC/DC converter: (i) an ultra-low-power boost converter (V$_{BAT}$) 
and (ii) a nano-power buck converter (V$_{buck}$) that 
can support up to 110mA output current. The BQ25570 has a programmable power good output signal (V$_{BAT\_ok}$) that indicates when the super-capacitor reaches a user-set voltage level. We set this signal to 2.1V. 
The V$_{BAT}$ converter 
is highly efficient when the storage element voltage level is above 1.8V but it is inefficient under this threshold (i.e. `cold-start'). The cold-start can be frequent if energy availability is low and can occur when adding or moving sensor nodes. 
In Figure \ref{fig:ChargeColdStart}-left, we show the charging time of the 3 super-capacitors sizes 
using V$_{BAT}$ under 750lux. It takes 1 day for the 220mF SC to charge from 0 to 3.6V 
and almost the entire charging-time is spent to exit cold-start operations (0 to 1.6V). 
The second DC-DC converter (V$_{buck}$) charges the super-capacitor at a much faster rate and the 1F element exits cold-start operations after 2.2 hours at 750 lux (Figure \ref{fig:ChargeColdStart}-right). 
However, V$_{buck}$ is more energy consuming after cold-start operations are over. Hence, we switch between the two charging modes using the circuit in Figure \ref{fig:CircuitMOS}-right. 
When the super-capacitor voltage level is low, the EMB raises V$_{buck}$ and V$_{BAT}$. Since the V$_{BAT\_ok}$ signal is low, the inverter (P-2 and N-3) blocks P-1 while N-2 is conducing and charges the super-capacitor through V$_{buck}$. Using a gate-source voltage threshold of 1.7V, N-2 charges the super-capacitor till 2.1V by setting V$_{Buck}$ to 3.8V. 
At this point, the V$_{BAT\_ok}$ signal turns on and connects V$_{BAT}$ to the storage element through P-1. 
Finally, N-1 powers the MCU once the super-capacitor reaches 2.1V (Figure~\ref{fig:CircuitMOS}-left). 
\vspace{-1mm}
\subsubsection{Wireless Communication Protocol}
We use Bluetooth Low Energy (BLE) as it offers advantages for indoor environments~\cite{paper:Cosero}:
\textit{(i)} BLE is more energy efficient w.r.t other technologies (e.g. ZigBee) in transmitting small data packets called advertisements. 
\textit{(ii)} advertisements can be used to perform indoor occupancy detection~\cite{paper:Cosero}.

\vspace{-2mm}
\begin{figure}
  \centering
     \includegraphics[width=0.5\linewidth]{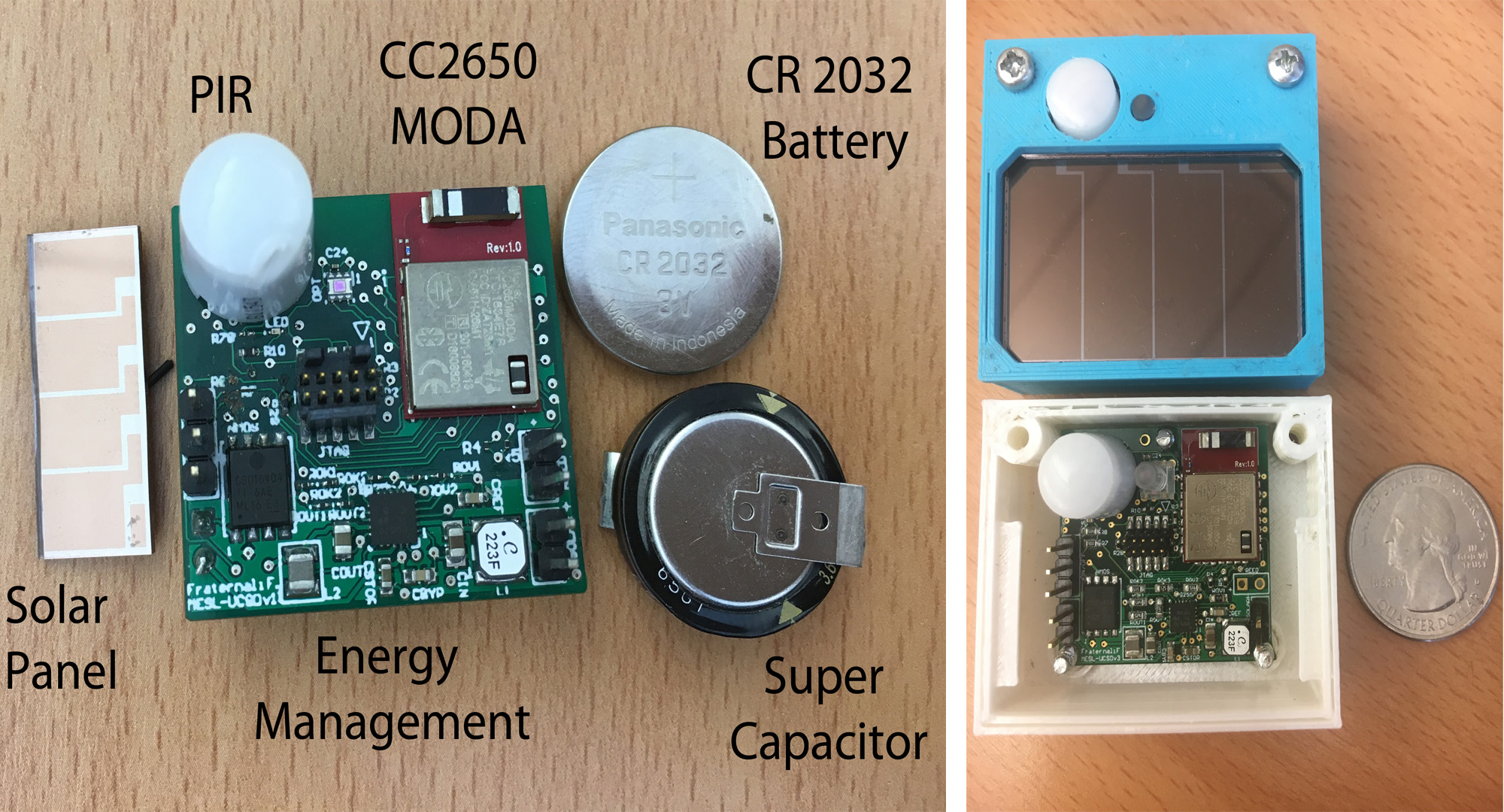}
    \vspace{-2mm}
    \caption{Pible: a Perpetual Indoor BLE Mote} 
	\label{fig:Pible}
	\vspace{-2mm}
\end{figure}

\vspace{-1mm}
\subsection{Wireless Sensor Network Architecture}
Pible-nodes send a data packet to the closest Base Station (BS) that stores and sends the data to the cloud for post-processing. 
We used a Raspberry PI equipped with a BLE USB dongle as a Base Station. To monitor the nodes' status, each Pible node sends to the BS information such as sensor data, QoS state, and voltage level.
 \vspace{-2mm}

\section{Power Management Algorithm}
\label{sec:Algorithm}
A mote with a 1F super-capacitor 
that advertises every 100ms 
lasts 1.9 hours without energy harvesting. Our algorithm dynamically changes the rate of advertising depending on energy availability to increase lifetime. 
It uses a simple sensor specific lookup table, and a lighting availability prediction to set the sensing rate. All the algorithmic decisions are made inside Pible and are the following: 

\begin{sloppypar}
\textbf{Setting the Sensing Quality}:
We use the super-capacitor voltage level to make a coarse adaptation of the QoS to use. We divide the usable voltage level (from 3.6V to 2.1V) into 7 states to maintain MCU memory requirements low.
Table \ref{tab:QoS} shows the relation between the voltage level and QoS selected for the sensors in Pible. 
The different sensing periods are manually assigned to different QoS levels based on the power measurements in Table \ref{tab:Energy} and empirical observations and requirements given in indoor sensing literature \cite{paper:campbell_2}. 
Event-driven sensors need to be continually operating to send a packet as an event occurs. To save power, we turn off the sensors for a fixed period as soon as an event occurs. The longer the switched off period, the more likely an event will be missed, but it will also save more power. 
For advertising applications, BLE indoor localization systems exploit advertisement rates between 0.1s to 0.9s~\cite{link:advrate}. We assign QoS levels 
to best meet this requirement.
\end{sloppypar}

\textbf{Light Intensity Prediction}:
Prediction of light intensity is used to refine the next QoS. The system stores and compares the last 5 light intensity levels: if the light read is close to 0 (i.e. light off) or decreasing, the algorithm decreases the QoS state while if the light intensity is increasing the QoS state is increased.

\textbf{Super-Cap Voltage Level Prediction}:
we store the last 5 voltage values of the super-cap and if the voltage level decreases or remains stable over time, the algorithm lowers the QoS by a level while if it is increasing the system increases the QoS.
5 levels of voltage and light intensity help us capture the short term trends while keeping the MCU compute and memory requirements low.
Between two sensing intervals, the system goes to sleep.


\begin{table}
\footnotesize
\centering
  \caption{Relation between Voltage Level and QoS to enable applications requiring sensing and user-position.}
  \label{tab:QoS}
  \vspace{-4mm}
  \begin{tabular}{cccccc}
    \toprule
    QoS  & Voltage & QoS & QoS-PIR & QoS\\
    State & [V] & Sensings [s]& Detection [s] & Advertising [s] \\
    \midrule
    7 & 3.6 - 3.4& 20 & 10 & 0.1\\
    6 & 3.4 - 3.2& 40 & 20 &0.2\\
    5 & 3.2 - 3.0& 60 & 30 &0.4\\
    4 & 3.0 - 2.8& 120 & 60& 0.64\\
    3 & 2.8 - 2.6& 240 &120& 0.9\\
    2 & 2.6 - 2.4& 300 &300& 2\\
    1 & 2.4 - 2.1& 600 &600& 5\\
  \bottomrule
  \vspace{-7mm}
\end{tabular}
\end{table}

\vspace{-2mm}

\section{Experimental Results}
\label{sec:Experimental}
We report experimental results of Pible motes deployed in a real building. By avoiding cold start-operation (e.g. Figure \ref{fig:CircuitMOS}), we use a 1F super-capacitor for all our experiments.

\vspace{-2mm}
\subsection{Pible in the Wild}
\label{wild}
We first describe the specifics of the sensors used.

\textbf{1 Sensor}: 
We configured Pible for sensing pressure. 
\textbf{5 Sensors}:
We configured Pible for sensing 5 sensors: light, ambient-temperature, object-temperature, pressure and humidity. 
\textbf{PIR Detection}:
 To compare its performance, we placed a battery powered PIR node near Pible. 
\textbf{Broadcasting BLE Advertisements}:
To measure the performance, an external Base Station queries the nodes for light, QoS State, and super-cap voltage levels every 10 minutes.

Table \ref{tab:light-energy} quantifies the average-QoS per day achieved by placing 20 Pible-nodes on 5 different indoor locations. 
For each node, we collect data for > 15 days.
\begin{table}[ht]
\centering
\footnotesize
\vspace{-3mm}
\caption{Position-Application-QoS Trade-Off for  15 Pible-nodes. Data are averaged per day.}
\vspace{-3mm}
\label{tab:light-energy}
\begin{tabular}{c|cc|cc|cc|cc}
\toprule
    	\small{} &  \multicolumn{2}{c }{1 Sensor} &  \multicolumn{2}{c }{5 Sensors} & \multicolumn{2}{ c }{Advertising} & \multicolumn{2}{ c }{PIR} \\ 
	\cline{2-9}
	\small{}  & \small{Light} & \small{QoS} & \small{Light} & \small{QoS} & \small{Light} & \small{QoS}& \small{Light} & \small{Event[\%]}\\
	\small{}  & \small{[lux]} & \small{[s]} & \small{[lux]} & \small{[s]}& \small{[lux]} & \small{[s]}& \small{[lux]} & \small{Detect}\\ 	\midrule
	Door & \small{121} & \small{337}& \small{112} & \small{564}  & \small{175} & \small{1.9} & \small{116}  & \small{71}\\ \hline
    Center   & \small{246} & \small{128}& \small{227} & \small{251}& \small{312} & \small{0.9} & \small{395} & \small{94}\\ 
    Office  & \small{} & \small{}& \small{} & \small{}& \small{} & \small{} & \small{} & \small{}\\ \hline
    window& \small{7k} & \small{79}& \small{9k} & \small{95}& \small{6k} & \small{1.3} & \small{8k}  & \small{87} \\ \hline
    Stairs  & \small{235} & \small{94}& \small{238} & \small{160}& \small{241} & \small{0.6} & \small{} & \small{32} \\ 
    Access & \small{} & \small{}& \small{} & \small{}& \small{} & \small{} & \small{} & \small{} \\ \hline
    Confer & \small{1k} & \small{75}& \small{427} & \small{286} & \small{1k} & \small{0.8} & \small{1k} & \small{97}\\
    Room  & \small{} & \small{}& \small{} & \small{} & \small{} & \small{} & \small{} & \small{}\\
    \bottomrule
    \end{tabular}
    \vspace{-3mm}
\end{table}
The application executed and the placement of the node play a fundamental role on the QoS achieved. 
While nearby a window, Pible sends 5 sensor data every 95 seconds on average and 1 sensor data every 79 seconds. 
The \textit{Door} case has an average light of 112 lux and sends 5 sensors data every 564s.

\begin{figure}[ht]
	\includegraphics[width=\linewidth]{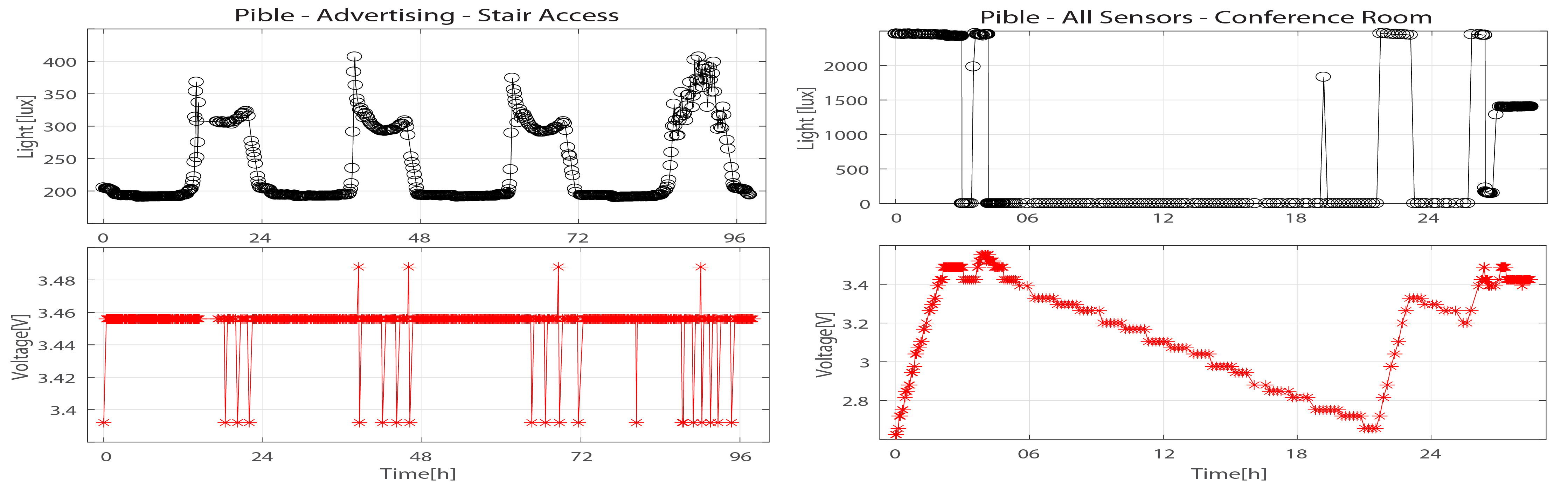}
	\vspace{-5mm}
	\caption{Pible in the Wild: (left) Advertising in a Stair Access Building Hall, (right) 5 Sensors in a Conference Room}
	\label{fig:Real2}
	\vspace{-5mm}
\end{figure}

Figure \ref{fig:Real2}-left shows Pible used for BLE advertising and placed in a stairway where internal lights are always on for security reasons. Even if the average-light per day is similar to other locations (i.e. \textit{center of office}), the QoS is better due to consistent light availability. The power management algorithm adapts the QoS by fully-charging the super-capacitor and then by increasing the QoS up to the maximum allowed. This validates the usefulness of the lookup table. 
Figure \ref{fig:Real2}-right shows results of Pible placed in a conference room while sensing 5 sensors. There are no windows in the room, and 
due to the intermittent light availability due to the presence of people, we place the node close to an internal light (i.e. 2500 lux) otherwise, the node would not have sustained continual operations. 
Applications such as advertising and one sensing perform well even in the presence of low light (Table \ref{tab:light-energy}). 
For PIR event detection, 
the placement of Pible in the center of an office or in a conference room detects occupancy events with an accuracy respectively of 94\% and 97\%. In these conditions, the light available is adequate to recharge the super-capacitors and most events are detected. 
Stair access case achieves only 32\% since  
too many people pass through the area and time between events is not enough to recharge the super-capacitor.
\vspace{-2mm}
\subsection{Comparison with State of the Art Solutions}
\label{Piblecompare}
The QoS, depends on the application - for periodical sensing sensors (light) it is to sensing rate, for event-driven sensors (PIR) it is the success rate of event capture.
We compare the QoS achieved by Pible w.r.t. battery powered systems~\cite{paper:yuvraj_1} and a pure solar energy harvesting powered architecture (PEH) for buildings~\cite{paper:campbell_2}, since 
they represent the extremes of our design space. We consider the location \textit{Center of Room} and emulate the performance of the baselines.
\begin{table}[ht]
\footnotesize
\centering
\vspace{-2mm}
\caption{Pible QoS Comparison with State of Art Solutions: Battery System end Pure Energy Harvesting Architecture}
\vspace{-2mm}
\label{tab:QoS_comparison}
\begin{tabular}{cccccc}
\toprule
    System 		& 	QoS [s] 	&  QoS [s] & PIR Events & QoS [s] & Working\\ 
    	& 	1 Sensor 	&  5 Sensor & Detect [\%] & Advertise & Operations\\ 
\midrule
	Battery & 60				 & 		60		  & 100 &		0.1 to 0.9	& 	few months\\
	Pure EH \cite{paper:campbell_2}		& 139 				& 	309			  &	96	&	 NA					& 	when light-on\\
	Pible									& 128				 & 		251		  & 94	&	0.6 to 0.8		& Perpetual\\
    \bottomrule
    \end{tabular}
    \vspace{-2mm}
\end{table}

\vspace{-1mm}
\subsubsection{Quality of Service}
\label{Quality_of_service}

\textbf{Sensing 1 Sensor:} PEH~\cite{paper:campbell_2} uses a 1.56mJ storage capacitor to 
sense and send a sensor data without any standard communication protocol. 
Pible needs 3.20 mJ to sense and send a data using BLE. Since Pible uses the light sensor to monitor the node, we simulate a PEH system that sense and send 2 sensor data (i.e. 6.40mJ). 
Furthermore, we compare Pible to a battery system from MicroDAQ that sends sensor data every minute using BLE. Results are reported on Table \ref{tab:QoS_comparison}. 
The PEH sends data with an average of 139s and Pible 128s. The results are better for Pible since the PEH stops sending data when the light is off.
\textbf{Sensing 5 Sensors:}
In this case, the PEH needs to accumulate 8mJ for the 5 sensors. 
Table \ref{tab:QoS_comparison} shows that a PEH can send data with an average per day of 309s while Pible of 251s. 
\textbf{PIR Detection:}
A PIR detection together with a BLE transmission of light and QoS requires 5.12mJ. 
By looking at the time-stamp and data given by the battery system, 96\% of the events were captured. PEH provides a good accuracy in detection since most of the events happen when light is on.
\textbf{Occupancy Detection:} BLE state of the art localization systems exploit advertisement rates between 0.1s and 0.9s \cite{link:advrate}.
Table \ref{tab:light-energy} shows that the average QoS per day spans between 0.6 and 0.9 when Pible is placed in a Center of Office, Windows or Conference Room. 
We posit that occupancy based triggers can tolerate latencies of up to 1 second and the Pible QoS is acceptable for these applications.

\vspace{-2mm}
\subsection{Limitations}
\label{limitations}
\vspace{-1mm}
\subsubsection{Manual creation of lookup table and thresholds:} we manually write the QoS for each sensing application given the super-capacitor voltage level (i.e. Table \ref{tab:QoS}). 
As a future work, we will explore machine learning methods to automatically configure the sensors to different lighting and application demands.

\vspace{-2mm}
\subsubsection{Operations with no light} 
With no light and by using the lowest QoS, Pible maintains operations for 19 hours by advertising, 27 hours for 5 sensing applications and 31 hours for 1 sensor applications. During our 15-days experiment, we were always able to achieve perpetual operations since presence of people was constant. 
If light is not present for more than those times, the node stops working. This can be avoided by (i) moving Pible to a closer source of light; (ii) using a bigger super-capacitor or a bigger solar panel. 

\vspace{-2mm}


\section{Conclusion and Future Work}
\label{sec:Conclusion}
We presented Pible: a battery-free mote for Perpetual Indoor BLE applications build out of commercial off the shelf components. 
Pible leverages ambient light and a power management algorithm to maximize the quality of service while satisfying perpetual working operations. We tested Pible in a real-world environment, showing that Pible maintained continual operations for 15 days and for 5 different lighting conditions. As a future work, we will explore different prediction mechanisms.
\vspace{-2mm}

\section*{Acknowledgments}
\vspace{-1mm}
This work is supported by the National Science Foundation grants CSR-1526237, TWC-1564009 and BD Spokes 1636879

\vspace{-2mm}

\bibliographystyle{ACM-Reference-Format}
\renewcommand*{\bibfont}{\footnotesize}
\bibliography{BuildSys-2018}

\end{document}